\documentclass[12pt]{article}
\usepackage{longtable}
\usepackage{graphicx}
\usepackage{lscape}
\usepackage{rotating}
\usepackage{mathtools}
\usepackage[section]{placeins}
\usepackage[top=1in, left=1in, right=1in, bottom=1in]{geometry}
\usepackage[parfill]{parskip}	
\usepackage{graphicx}
\usepackage{caption}
\usepackage{subcaption}
\usepackage[export]{adjustbox}
\usepackage{amssymb}
\usepackage{array}
\usepackage{arydshln}
\usepackage{epstopdf}
\usepackage{bbm}
\usepackage{bm}
\usepackage{amssymb}
\usepackage{amsmath}
\usepackage{amsfonts}

\usepackage{lscape}
\usepackage{enumitem}
\usepackage{rotating}
\usepackage{setspace}
\usepackage{threeparttable}
\usepackage{booktabs}
\usepackage[compact]{titlesec}
\usepackage{lmodern}
\fontfamily{lmtt}\selectfont
\usepackage[T1]{fontenc}

\usepackage[nocitation]{apacite}
\usepackage{natbib}
\bibpunct{(}{)}{;}{a}{}{,}
\usepackage{hyperref}
\usepackage{titling}
\usepackage{pifont}
\usepackage[capposition=top]{floatrow}
\newcommand{\subtitle}[1]{%
  \posttitle{%
    \par\end{center}
    \begin{center}\large#1\end{center}
    \vskip0.5em}%
}
\usepackage{amsthm}
\usepackage{amsmath} 
\usepackage{multirow}

\usepackage{multicol}
\usepackage{mathrsfs}

\usepackage{mathrsfs}

\usepackage{color}

\usepackage{array}
\newcolumntype{C}[1]{>{\centering\arraybackslash}m{#1}}
\begin{document}
\pagestyle{plain}

\newtheoremstyle{mystyle}
{\topsep}
{\topsep}
{\it}
{}
{\bf}
{.}
{.5em}
{}
\theoremstyle{mystyle}
\newtheorem{assumptionex}{Assumption}
\newenvironment{assumption}
  {\pushQED{\qed}\renewcommand{\qedsymbol}{}\assumptionex}
  {\popQED\endassumptionex}
\newtheorem{assumptionexp}{Assumption}
\newenvironment{assumptionp}
  {\pushQED{\qed}\renewcommand{\qedsymbol}{}\assumptionexp}
  {\popQED\endassumptionexp}
\renewcommand{\theassumptionexp}{\arabic{assumptionexp}$'$}

\newtheorem{assumptionexpp}{Assumption}
\newenvironment{assumptionpp}
  {\pushQED{\qed}\renewcommand{\qedsymbol}{}\assumptionexpp}
  {\popQED\endassumptionexpp}
\renewcommand{\theassumptionexpp}{\arabic{assumptionexpp}$''$}

\newtheorem{assumptionexppp}{Assumption}
\newenvironment{assumptionppp}
  {\pushQED{\qed}\renewcommand{\qedsymbol}{}\assumptionexppp}
  {\popQED\endassumptionexppp}
\renewcommand{\theassumptionexppp}{\arabic{assumptionexppp}$'''$}

\renewcommand{\arraystretch}{1.3}

\newcommand{\argmin}{\mathop{\mathrm{argmin}}}
\makeatletter
\newcommand{\grande}{\bBigg@{2.25}}
\newcommand{\enorme}{\bBigg@{5}}

\newcommand{\blind}{0}

\newcommand{\tit}{Randomized and Balanced Allocation of Units \\ into Treatment Groups Using the \\ Finite Selection Model for R}

\if0\blind

{\title{\tit}
\author{Ambarish Chattopadhyay\thanks{Department of Statistics, Harvard University, 1 Oxford Street
Cambridge, MA 02138; email: \url{ambarish_chattopadhyay@g.harvard.edu}.} \and Carl N. Morris\thanks{Department of Statistics, Harvard University, 1 Oxford Street
Cambridge, MA 02138; email: \url{carl.morris@comcast.net}.} \and Jos\'{e} R. Zubizarreta\thanks{Departments of Health Care Policy, Biostatistics, and Statistics, Harvard University, 180 Longwood Avenue, Office 307-D, Boston, MA 02115; email: \url{zubizarreta@hcp.med.harvard.edu}.}
}

\date{} 

\maketitle
}\fi

\if1\blind
\title{\bf \tit}
\date{} 
\maketitle
\fi

\begin{abstract}
\noindent The original Finite Selection Model (FSM) was developed in the 1970s to enhance the design of the RAND Health Insurance Experiment (HIE;  \citealt{newhouse1993free}).
At the time of its development by Carl Morris \citep{morris1979finite}, there were fundamental computational limitations to make the method widely available for practitioners.
Today, as randomized experiments increasingly become more common, there is a need for implementing experimental designs that are randomized, balanced, robust, and easily applicable to several treatment groups. 
To help address this problem, we revisit the original FSM under the potential outcome framework for causal inference and provide its first readily available software implementation.
In this paper, we provide an introduction to the FSM and a step-by-step guide for its use in R.
\end{abstract}


\begin{center}
\noindent Keywords:
{causal inference, covariate balance, experimental design, randomization, R}
\end{center}
\clearpage
\doublespacing

\singlespacing
\pagebreak
\tableofcontents
\pagebreak
\doublespacing


\section{Introduction}
\label{sec_introduction}

\subsection{Origin of the FSM}
\label{sec_origin}

The original Finite Selection Model (FSM) was proposed and developed at the RAND Corporation in the 1970s to enhance the experimental design for the now famous Health Insurance Experiment (HIE), the second of several large-scale national public policy experiments of that era.  
Among others, the experimental findings from the HIE helped to understand the consequences of health care financing on people's health and supported the reorganization of private health insurance in the US (see \citealt{newhouse1993free}). 

The FSM procedure was conceived and developed by Carl Morris \citep{morris1979finite}. The FSM generalizes a process familiar to youngsters world-wide who have played schoolyard games, when two teams are assembled from a group of willing participants by having team captains take turns, at each opportunity choosing the best remaining player for his or her team.  It is intuitive that if the two captains are equally well-informed (or even equally poorly informed) about the abilities of the available individuals, the resulting teams are likely to be well ``matched'', or well ``balanced'' in the language of experimental design.  If the captains are also equally well-informed strategically about winning the contest, they will assemble teams that are likely to compete at a higher level.

In the FSM acronym, ``Finite'' recognizes the finiteness of the list of available units (or subjects) being chosen, as opposed to design tools that make choices from a possibly infinite list of units.  ``Selection'' emphasizes that treatment ``captains'' (or ``choosers'') select units for treatments, taking turns in a fair and random order. In the FSM, randomness is an integral part of the treatment allocation process, providing a basis for inference.  More specifically, the order in which treatments select units is determined by a randomized ``selection order matrix.'' At each turn, the choosing treatment adds the single remaining unit that maximally improves the combined quality of its current group of units. 

In the HIE, the FSM involved having 13 treatments (insurance plans) choose families in a balanced and controlled random order. This was done separately for each of the HIE’s six national experimental sites (four cities and two counties in the USA).  In each site, a group of eligible families had been identified, each willing to participate by accepting which ever treatment it was offered.  Selections were based on each family's known vector of 23 covariates chosen to determine the family's value for predicting health utilization regressions.

A primary advantage of using the FSM as part of an experimental design lies in its ability to improve covariate balance for a large list of baseline covariates (23 in the HIE) across a large number of treatment groups (13 in the HIE). Other FSM advantages lie in not needing to dichotomize continuous variables.  For an overview of the FSM and how it was used in the HIE (see \citealt{morris1993the}).  

\subsection{The FSM package}
\label{sec_package_intro}

At the time of the conceptual development of the original FSM, there were fundamental computational limitations to making the method widely available for practitioners. However, today computation is no longer a binding constraint, and as randomized experiments are becoming ever more common, there is a need for implementing experimental designs that are randomized, balanced, robust, and easily applicable to multiple treatment groups. To help address this problem, we revisit the original FSM under the potential outcome framework for causal inference and provide its first readily available software implementation.

In this paper, we provide an introduction to the FSM and illustrate how to use it in the new \texttt{FSM} package for R. In Section \ref{sec_theory}, we delineate the framework and notations, explain the motivating idea behind the FSM, and describe its main components.
In this section we also discuss how to perform diagnostics and conduct inference after allocation with the FSM.
In Section \ref{sec_stepbystep}, we illustrate how to use the \texttt{FSM} package with step-by-step code examples.
Finally, in Section \ref{sec_discussion}, we comment on further software developments for the FSM.
A complete R script to replicate the main analyses in this paper and a detailed description of the main functions in the \texttt{FSM} package can be found in the Appendix.
We ask the reader to refer to the package documentation for detailed descriptions of other, more specialized functions in the package.
\section{Designing an experiment using the FSM}
\label{sec_theory}

\subsection{Framework}
\label{sec_framework}

Consider a sample of $N$ units indexed by $i = 1, 2, ...,N$.
Our goal is to randomly assign the units into $g$ treatment groups, $T_1$, $T_2$, ..., $T_g$, of sizes $n_1, n_2,..., n_g$, respectively, where $n_1 + n_2 + ... + n_g  = N$. Let $\bm{X}_i$ be the $k \times 1$ vector of baseline characteristics or covariates of unit $i \in \{1,2,...,N\}$. 
Write $\bm{Z} = (Z_1, Z_2, ..., Z_N)^\top$ as the $N \times 1$ vector of treatment group labels, with $Z_i = j$ if unit $i$ is assigned to treatment group $j \in \{1,2,...,g\}$. 

We base our discussion on the potential outcome framework for causal inference \citep{neyman1923application, rubin1974estimating, imbens2015causal}. 
Under the Stable Unit Treatment Value Assumption (SUTVA; \citealt{rubin1980randomization}), we write the potential outcome of unit $i$ under treatment level $j$ as $Y_i(j)$.
The unit level causal effect of treatment $j'$ relative to treatment $j''$ for unit $i$ is given by $Y_i(j')- Y_i(j'')$.

Throughout this paper, we use as a running example the Nationally Supported Work (NSW) experimental data set by Lalonde (\citeyear{lalonde1986evaluating}; see also \citealt{dehejia1999causal}), or simply, the Lalonde data set. 
The experiment evaluates the impact of the NSW program on earnings. The complete study consists of $185$ treated individuals or units (enrolled in the training program), $260$ control units (not enrolled in the training program).\footnote{This data set can be downloaded from \url{https://users.nber.org/~rdehejia/nswdata.html}} There are eight baseline covariates, namely, \texttt{Age} (age measured in years), \texttt{Education} (years of education), \texttt{Black} (indicator for Black race), \texttt{Hispanic} (indicator for Hispanic race), \texttt{Married} (indicator for married status), \texttt{Nodegree} (indicator for high school dropout), \texttt{Re74} (earnings in 1974), and \texttt{Re75} (earnings in 1975). For ease of exposition, in the remainder of this section we consider a reduced version of the data set consisting of a randomly chosen subset of $N=12$ units and $k=1$ covariates \texttt{Age}, as described in Table \ref{data_lalonde_short}.

\begin{table}[!ht]
\centering
\scalebox{0.85}{
\begin{tabular}{cc} 
\toprule
Index & Age\\ 
  \hline
1 & 19\\ 
  2 & 20\\ 
  3 & 20\\ 
  4 & 20\\ 
  5 & 23\\ 
  6 & 24\\ 
  7 & 24\\ 
  8 & 25\\ 
  9 & 25\\ 
  10 & 28\\ 
  11 & 31\\ 
  12 & 41\\ 
  \hline
  Mean & 25\\
   \bottomrule
\end{tabular}
}
\caption{\footnotesize A sample of 12 units selected from the Lalonde data set. In this abridged version of the data set, we consider a single covariate, \texttt{Age}.} 
\label{data_lalonde_short}
\end{table}


\subsection{Motivating the FSM: wisdom from schoolyard games}
\label{sec_games}

The FSM can be motivated by the familiar situation of building schoolyard teams from a given pool of players. In a game with two teams, the team captains often select the players for their respective teams by taking turns. After every stage of selection, the number of players available for selection is reduced by one. Typically, when their turn a choosing captain selects the best player still available to improve their team. This process of selection continues until all players are chosen. In the FSM, the treatment groups play the role of the team captains and the units play the role of the players. The FSM provides a method of assignment where the treatment groups choose the units, instead of the units randomly choosing a treatment group.

In the schoolyard game example, suppose that both captains are equally knowledgeable of what constitutes a strong team and of the players’ abilities.
Then, if the captains take turns in a fair order, the resulting two teams are expected to be of similar strength. 
Moreover, if the captains are well-informed about the players' abilities, the two individual teams are likely to be reasonably strong. 
Interestingly, even if the captains are misinformed about players' abilities, they are equally misinformed and hence the two teams are still expected to be equally strong. Intuitively, this suggests that with a proper specification of the order and the criteria of selection, the FSM can yield groups that are well-balanced along with high degree of efficiency and robustness. Finally, this idea of selection of units by treatments immediately applies to the case with more than two treatment groups.

\subsection{Components of the FSM}
\label{sec_components}

The FSM has two main components, the selection order matrix (SOM) and the selection function, which we describe as follows.

\subsubsection{Selection order matrix (SOM)}
\label{sec_som}

The selection order matrix (SOM) is a matrix specifying the order in which the treatment groups select the units. The matrix usually has two columns, the first column indicating the stage of selection and the second column indicating the label of the treatment group that gets to choose at that stage. In the case of a stratified design, the SOM may include a third column of stratum labels. A good SOM should be fair so that the order of choices made by one treatment group does not unjustly deprive its opponents of important units. In addition, an SOM should be random, since randomization helps protect against imbalances due to unobserved covariates and at the same time provides a basis for inference. 

Given the values of $N$, $g$, and the group sizes $n_1,...,n_g$, one can use different methods to generate an SOM (see the Appendix for a list of options available in the \texttt{FSM} package). For two treatment groups, we recommend using the Sequentially Controlled Markovian Random Sampling (SCOMARS) algorithm \citep{morris1979finite, morris1983sequentially} to generate an SOM. In this setting, SCOMARS provides a vector of conditional probabilities whose $j$th element is equal to the probability that one of the treatments groups, say, $T_2$, gets to choose at the $j$th stage, conditional on the number of choices made by $T_2$ upto the $(j-1)$th stage ($j \in \{1,2,...,N\}$). SCOMARS ensures that, if $T_2$ has gotten more (respectively, less) than its fair share of choices up to a certain stage, it has a low (respectively, high) probability of making a choice in the next stage. The \texttt{FSM} package has a built-in function for SCOMARS (see Section \ref{sec_choosesom} and the Appendix for details). 
In our example, consider the problem of allocating the 12 units in Table \ref{data_lalonde_short} into two groups of equal size. One instance of an SOM generated using SCOMARS for this assignment problem is given in Table \ref{table_som1}.

\begin{table}[!ht]
\centering
\scalebox{0.85}{
\begin{tabular}{ccc}
  \toprule
Stage of selection & Prob($T_2$ selects) & Treatment \\ 
  \hline
1 & 0.5 & 2 \\ 
  2 & 0.0 & 1 \\ \hdashline
  3 & 0.5 & 1 \\ 
  4 & 1.0 & 2 \\ \hdashline
  5 & 0.5 & 1 \\ 
  6 & 1.0 & 2 \\ \hdashline
  7 & 0.5 & 1 \\ 
  8 & 1.0 & 2 \\ \hdashline
  9 & 0.5 & 1 \\ 
  10 & 1.0 & 2 \\ \hdashline
  11 & 0.5 & 2 \\ 
  12 & 0.0 & 1 \\ 
   \bottomrule
\end{tabular}
}
\caption{\footnotesize An SOM generated using SCOMARS with $N = 12$ units and $g= 2$ groups of equal sizes.}
\label{table_som1}
\end{table}

 In this example, the vector of conditional probabilities are given in column 2 of Table \ref{table_som1}. $T_2$, which has a 0.5 probability of being the chooser in the first stage of selection, ends up being the chooser. Given that $T_2$ gets to select in the first stage, SCOMARS forces $T_1$ to select in the second stage. The same process continues for the following pairs of stages, independently of the previous pairs. Therefore, in the case of two treatment groups with equal sizes, SCOMARS is equivalent to randomly permuting the treatment labels $(1,2)$ independently across successive pairs of stages, yielding a random SOM. Since no treatment group is preferred over the other, an SOM generated in this way is fair. 

Although SCOMARS is designed only for two groups, for multiple treatment groups of equal sizes we can generate a fair and random SOM using a similar permutation-based algorithm. More formally, for assigning $N = cg$ units to $g$ treatment groups (where $c$ is some positive integer), an SOM can be created by successively generating $c$ independent random permutations of the $g$-tuple $(1,2,...,g)$.
In the case of multiple groups with unequal sizes, one can use several strategies to generate an SOM. One such possibility is to first combine the groups appropriately into two groups and then apply SCOMARS repeatedly to split the combined groups into the desired number of groups. For example, with $n_1 = 2$, $n_2 = 4$, $n_3 = 6$, we can first apply SCOMARS to split the 12 units to the groups $(T_1, T_2)$ and $T_3$ of sizes $n^*_1 = 2+4 =6$ and $n_3 = 6$ respectively. Subsequently, we can use SCOMARS again to split the combined group $(T_1, T_2)$ into two groups of sizes $n_1 = 2$ and $n_2 = 4$.


\subsubsection{Selection function}
\label{sec_selfunc}

The second component of the FSM is called the selection function, which provides the common criteria by which the treatment groups select units. In the schoolyard games example, a team's selection function determines who is the best remaining player at every stage of selection, conditional on the choices it already has made. In principle, one can use any criterion as the selection function (see the Appendix for a list of some available options in the \texttt{FSM} package). A robust choice is the \textit{D-optimal} selection function. This selection function implicitly considers a linear model $Y_i(j) = \bm{\beta}^\top_j (1,\bm{X}^\top_i)^\top + \epsilon_{ij}$ of the potential outcome under treatment level $j \in \{1,2,...,g\}$, where $\mathbb{E}[\epsilon_{ij}|\bm{X}_i] = 0$.\footnote{ One can also consider a linear model without the intercept term.} A D-optimal design minimizes the generalized variance of the ordinary least squares (OLS)\footnote{Or weighted least squares.} estimator of $\bm{\beta}_j$ based on the fitted model of $Y_i(j)$ in $T_j$, which is equivalent to maximizing $\text{det}(\underline{\bm{X}}^\top_j \underline{\bm{X}}_j)$, where $\underline{\bm{X}}_j$ is the design matrix based on all the units selected for $T_j$, $j \in \{1,2,...,g\}$. 

The D-optimal selection function targets this aspect of a D-optimal design, but in a sequential manner. To formalize, without loss of generality, suppose $T_1$ has the turn to choose at stage $r$ ($r\in \{1,2,...,N\}$). Let $\underline{\bm{X}}_{r-1,1}$ be the corresponding design matrix in $T_1$ based on the units $T_1$ has selected upto stage $r-1$. Using the D-optimal selection function, $T_1$ chooses the unit from the remaining pool of $N-(r-1)$ units that maximizes the determinant of $\underline{\bm{X}}_{r,1}^\top \underline{\bm{X}}_{r,1}$, where $\underline{\bm{X}}_{r,1}$ denotes the resulting design matrix for $T_1$ after stage $r$. Ties are resolved by randomly selecting one of the optimal units. In the special case of a single covariate, denoting $a$ as the covariate of a candidate unit and $\bar{X}_{r-1,1}$ as the mean of the covariate in $T_1$ up to the $(r-1)$th stage ($r \geq 2$), selection using the D-optimal selection function boils down to choosing the unit that maximizes $|a - \bar{X}_{r-1,1}|$ among all the units available for selection. When $r=1$, the treatment group attempts to choose the unit that maximizes $|a - \bar{X}_{\text{full}}|$, where $\bar{X}_{\text{full}}$ is the mean of the covariate in the full sample. 

Assuming a linear model of each potential outcome on \texttt{Age}, we illustrate this selection process using the data in Table \ref{data_lalonde_short} and the SOM in Table \ref{table_som1}. In Table \ref{assign1}, we augment the SOM in Table \ref{table_som1} with the columns of indices of the chosen units and their corresponding ages.\footnote{We remove the column of selection probabilites.}

\begin{table}[!ht]
\centering
\scalebox{0.85}{
\begin{tabular}{cccc}
  \toprule
Stage of selection & Treatment & Index & Age \\ 
  \hline
1 & 2 & 12 & 41 \\ 
  2 & 1 & 11 & 31 \\ \hdashline
  3 & 1 & 1 & 19 \\ 
  4 & 2 & 4 & 20 \\ \hdashline
  5 & 1 & 3 & 20 \\ 
  6 & 2 & 2 & 20 \\ \hdashline
  7 & 1 & 10 & 28 \\ 
  8 & 2 & 5 & 23 \\ \hdashline
  9 & 1 & 9 & 25 \\ 
  10 & 2 & 6 & 24 \\ \hdashline
  11 & 2 & 7 & 24 \\ 
  12 & 1 & 8 & 25 \\ 
   \bottomrule
\end{tabular}
}
\caption{\footnotesize Augmented SOM with columns corresponding to the indices of the selected units and their respective ages.}
\label{assign1}

\end{table}

The mean of \texttt{Age} in the full sample is 25 years. Since unit 12 is 41 years old, the farthest from the full sample mean in magnitude, it gets selected by $T_2$ in the first stage. In the second stage, unit 11 is 31 years old, the farthest in magnitude from 25 among the pool of 11 remaining units. Consequently, $T_1$ selects unit 11. In the third stage, $T_1$ tries to choose the unit whose age is farthest from $31$ and ends up choosing unit 1 with age 19. Similarly, $T_2$ tries to choose the unit whose age is farthest from $41$ and ends up choosing unit 4 with age 20, resolving a tie among units 2,3, and 4 randomly. This process continues until all the units are selected. The combination of an SOM and a choice of selection function allows us to generate a randomized assignment of units into the treatment groups. In this example, at the end of the selection process $T_1$ consists of 6 units with ages 31, 19, 20, 28, 25, 25, and $T_2$ consists of 6 units with ages 41, 20, 20, 23, 24, 24 (ages within each group are listed by the selection order).


\subsection{Design assessment}
\label{sec_designassess1}

In the design assessment stage, we evaluate the statistical properties of the FSM. 
We consider measures of covariate balance and relative efficiency of the design.
In causal inference, covariate balance is an intuitive objective to pursue in order to isolate the effect of the treatment from the confounding effects of imbalances in baseline covariates. 
More formally, covariate balance controls bias, but it also impacts variance. 
In most randomized designs, the act of randomization ensures that treatment groups are balanced \textit{on average}, i.e., over repeated realizations of the assignment mechanism. However, any given realization of the assignment mechanism can produce substantial imbalances. 
Such imbalances, in turn, can directly impact the variance of the average treatment effect estimator.

For instance, for $g=2$, consider fitting a linear outcome model with constant treatment effect $\tau$ (of treatment 1 versus treatment 2) and uncorrelated errors with constant variance $\sigma^2$. The \textit{Optimal Covariance Design (OPCODE)} theorem \citep{morris1993the} states that the model-based variance of the OLS estimator of $\tau$ is $\frac{\sigma^2}{ N s^2_{T_1} (1-R^2)}$, where $s^2_{T_1} = \frac{n_1}{N} (1 - \frac{n_1}{N})$ and $R^2$ is the square of the multiple correlation coefficient of the indicator of $T_1$ with the covariates; see also  \cite{greevy2004optimal}. The design that minimizes this variance (i.e., the OPCODE solution) satisfies $R^2 = 0$ (if feasible), which is equivalent to exact mean-balance of the covariates between the two treatment groups. However, under a misspecified outcome model, the previous OPCODE solution, even if feasible, may produce estimators with high variances. Hence covariate balance is a measure directly tied to the precision of the OLS regression-based effect estimator. Thus as part of the design assessment, balance checks help to evaluate the efficiency of a design under correct specification of the model as well as robustness of a design against model misspecifications.

In addition, given a collection of designs, one can compare the extent of information provided by each design using a measure of efficiency relative to a base design. Building on ideas from sample surveys \citep{kish1965survey}, the notion of relative efficiency can be translated into a notion of effective sample size (ESS). In the spirit of model- and randomization-based inference (described later in Section \ref{sec_inf1}), we propose notions of the model- and randomization-based ESS for a general collection of designs. Here the ESS of a design $D$ in a collection of designs is the number $n_{\textrm{eff}}$ that equates the full-sample size divided by the ratio of variance of the effect estimator under design $D$ to that under the design having the highest precision in the collection. 
By definition, for all designs in the collection, $n_{\text{eff}} \leq N$, and equality holds for the design having the highest precision in the collection. We recommend computing the ESS as it provides a more palpable measure of the total amount of information provided by a design, as compared to other standard measures of relative efficiency. 

As an example, we now evaluate the covariate balance and the ESS of the assignment obtained in Table \ref{assign1}. 
Table \ref{balance_small} includes the first two moments of \texttt{Age} for each treatment group and the corresponding means in the full sample. Table \ref{balance_small} indicates that the two treatment groups are reasonably balanced with respect to the means of \texttt{Age} and $\texttt{Age}^2$.

\begin{table}[!ht]
\scalebox{0.85}{
         \begin{tabular}{p{3cm}cc}
    \toprule 
    \multirow{2}{5cm}{Groups} & \multicolumn{2}{c}{Covariate transformations}\\
   \cline{2-3}
    & Age & $\text{Age}^2$\\
    \toprule
      Treatment 1  &  24.67  & 626.00\\
Treatment 2  &  25.33  & 693.67\\ \hline
Full sample  &  25.00  & 659.83\\
\bottomrule 
  \end{tabular}
}
       \caption{\footnotesize Means of \texttt{Age} and $\texttt{Age}^2$ for the assignment in Table \ref{assign1}, plus the corresponding values in the full sample.} 
\label{balance_small}       
\end{table}

To provide some perspective, we compare the overall degree of balance between the FSM and a completely randomized design (CRD). 
As a measure of imbalance, we compute the target absolute atandardized mean difference (TASMD, \citealt{chattopadhyay2020balancing}) of \texttt{Age} and $\text{\texttt{Age}}^2$ relative to their corresponding full-sample averages. In this context, the TASMD of a variable $U$ in $T_j$ relative to the full-sample is given by $\frac{|\bar{U}_j - \bar{U}|}{s_U}$. Here $\bar{U}_j$ and $\bar{U}$ are the means of $U$ in $T_j$ and the full-sample respectively, and $s_U$ is the standard deviation of $U$ in the full-sample. Smaller value of the TASMD indicates that $\bar{U}_j$ is closer to $\bar{U}$, implying better balance on $U$ in $T_j$ relative to the full-sample. In general, to measure imbalance on $U$ in $T_j$ relative to a target profile $U^*$ of interest, one can replace $\bar{U}$ by $U^*$ and $s_U$ by a suitable measure of standard deviation of $U$ (see \citealt{chattopadhyay2020balancing}).

For unweighted samples with two treatment groups, the TASMD of a covariate in a treatment group (relative to the full sample) is a scalar multiple of the more commonly used absolute standardized mean differences (ASMD) of that covariate. In fact, for two groups of equal sizes, the ASMD of a covariate can be shown to be no less than twice its TASMD in each of the two groups, relative to the full sample. Despite their equivalence in this context, for diagnostics we prefer using the TASMD to the ASMD for two main reasons. First, by construction, the TASMD not only quantifies the similarity (or lack thereof) in the vector of covariate-means of the treatment groups with respect to each other, but with respect to any target profile of interest. For instance, the target profile can be the vector of covariate-means of an arbitrary target population, or the vector of covariates of a target individual.
Thus, the TASMD allows judging the representativeness of each treatment group relative to the target, and in turn, assessing the transportability of the results of an experiment to the target population or individual of interest. Second, for weighted samples, TASMD is essential for balance diagnostics as it directly relates to the magnitude of the bias of the Haj\'{e}k estimator of treatment effect \citep{chattopadhyay2020balancing}.

Section \ref{sec_designassess2} displays a comparison of the TASMD for a realized assignment under the FSM to that under CRD. In this section, we focus on comparing the randomization distributions of the TASMD under CRD and the FSM. In Table \ref{crdversusfsm}, we show the mean and standard deviation of the TASMDs of \texttt{Age} and $\text{\texttt{Age}}^2$ in $T_1$, across 1000 randomizations of CRD and the FSM. Since the group sizes are equal, by symmetry, similar results hold for $T_2$.

\begin{table}[!ht]
\scalebox{0.85}{
         \begin{tabular}{p{3cm}cc}
    \toprule 
    \multirow{2}{5cm}{Covariate \\ transformations} & \multicolumn{2}{c}{Design}\\
   \cline{2-3}
    & CRD & FSM\\
    \toprule
      Age  &  0.24 (0.15)  & 0.07 (0.01)\\
$\text{Age}^2$  &  0.25 (0.13)  & 0.11 (0.01)\\
\bottomrule 
  \end{tabular}
}
       \caption{\footnotesize Mean and standard deviation (in parenthesis) of the TASMDs of \texttt{Age} and $\texttt{Age}^2$ across 1000 randomizations of CRD and the FSM.} 
\label{crdversusfsm}      
\end{table}

Table \ref{crdversusfsm} exhibits better balance on the first two moments of \texttt{Age} under the FSM as compared to CRD since the corresponding means of the TASMDs are uniformly smaller under the FSM, with substantially smaller standard deviations. We note that for this example, the FSM implicitly assumes potential outcome models that are linear in \texttt{Age}, thus leading to reasonable balance on the mean of \texttt{Age} by design. However, despite not directly accounting for $\text{\texttt{Age}}^2$, the FSM exhibits substantially better balance on the mean of $\text{\texttt{Age}}^2$ than CRD. 

For symmetrically distributed covariates, the D-optimal selection function tends to balance all even functions of the covariates. By balancing transformations of the covariates that are not included in the assumed model, the FSM achieves reasonable robustness against model misspecification. 

\begin{figure}[!ht]
     \centering
     \begin{subfigure}[b]{0.48\textwidth}
         \centering
         \includegraphics[width=\textwidth]{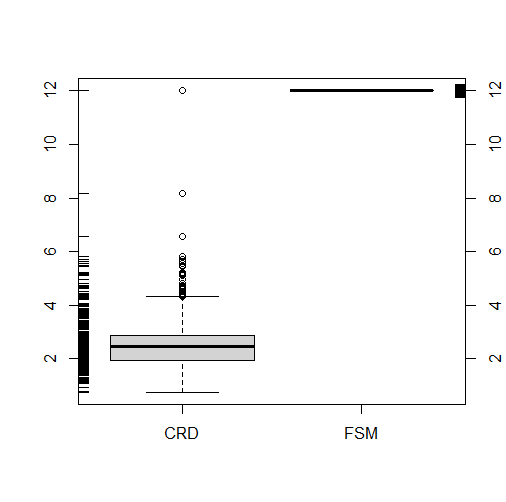}
         \caption{Linear in \texttt{Age}}
         \label{fig:linear age}
     \end{subfigure}
     \hfill
     \begin{subfigure}[b]{0.48\textwidth}
         \centering
         \includegraphics[width=\textwidth]{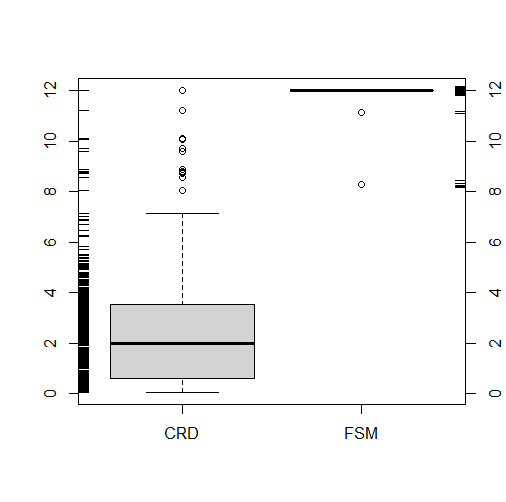}
         \caption{Quadratic in \texttt{Age}}
         \label{fig:quadratic age}
     \end{subfigure}
        \caption{\footnotesize Boxplots of the distribution of the model-based ESS for the given realization of the FSM with respect to 1000 independent realizations of CRD under model (a) (left) and model (b) (right).}
        \label{fig:ess boxplot}
\end{figure}

We now assess the efficiency of the given realization of the FSM relative to CRD by comparing the effective sample sizes (ESS). More specifically, given a simulated set of potential outcomes $\{ Y_i(1),Y_i(2): i = 1,2,...,N \}$, we generate $M$ independent assignments of CRD and compute the model-based ESS of the realized assignment of the FSM and the $m$th assignment of CRD ($m=1,2,...,M$). This yields a distribution of the model-based ESS for the realized assignment of the FSM relative to the randomization distribution of the corresponding model-based ESS of CRD. In this example we simulate the set of potential outcomes using two data generating processes, namely (a) $Y_i(1) = 30 - \bm{X}_i + \epsilon_i$, $Y_i(2) = Y_i(1)$ and (b) $Y_i(1) = -35.98 - \bm{X}_i + 0.1 \bm{X}^2_i +  \epsilon_i$, $Y_i(2) = Y_i(1)$, where $\epsilon_i$s are independent Normal random variables with mean $0$ and variance $16$. Model (a) is linear in \texttt{Age}, whereas model (b) is quadratic in \texttt{Age}. The coefficients in the two models are chosen such that the marginal means of $Y_i(1)$ are approximately the same under both models. The resulting boxplots of the distributions of the ESS for the FSM relative to $M=1000$ independent realizations of CRD are shown in Figure \ref{fig:ess boxplot}.

Under both scenarios, the given realization of the FSM outperforms CRD in terms of an ESS distribution that is substantially more concentrated towards the full-sample size of 12. In particular, Figure \ref{fig:linear age} shows that the ESS of the FSM is equal to 12 for every realization of CRD when the potential outcomes are generated under a linear model in \texttt{Age}. This is not surprising since the assumed linear potential outcome model coupled with the D-optimal selection function ensures that the model based variance estimates are reasonably small (see Section \ref{sec_selfunc}). However, even when the true model is quadratic in \texttt{Age}, the FSM almost always achieves the highest possible ESS (Figure \ref{fig:quadratic age}), suggesting once again that the FSM provides robustness against model misspecification. We see that in both the cases, the FSM effectively provides 5 or 6 times the information provided by a \textit{typical} CRD.

\subsection{Inference after the FSM}
\label{sec_inf1}

In randomized experiments, there are two common modes of inference: randomization- and model-based inference. Although randomization-based inference was Fisher's original proposal in the 1920s \citep{fisher1925statistical, fisher1935design, fisher1992statistical}, model-based inference has been the default mode of inference in the experimental design literature throughout much of the 20th century \citep{montgomery2001design}. 
However, over the last few decades, randomization-based inference has gained increasing attention under the potential outcome framework for causal inference (e.g., \citealt{rosenbaum2002observational, samii2012equivalencies, imbens2015causal}). 
Randomization-based inference explicitly takes into account the assignment mechanism that allocated the units into the treatments groups, often without resorting to large sample approximations. Here the sets of potential outcomes and covariates are viewed as fixed and the only source of randomness is the random assignment of treatments (see Chapter 2 of \citealt{rosenbaum2002observational} and chapters 5--7 of \citealt{imbens2015causal} for overviews). 
In model-based inference, the sample is assumed to be randomly drawn from a large super-population, rendering the set of potential outcomes and covariates random. Here, randomness stems from both the random sampling of units and the random assignment of treatments. The methodology and implementation of the FSM does not explicitly depend on which mode of inference we follow. However, the mode of inference may impact the confidence intervals obtained in the outcome analysis stage. 
The \texttt{FSM} package includes functions to perform both randomization- and model-based inference for average treatment effects. 
In this paper, we focus on model-based inference.

In model-based inference, a common causal estimand of interest is the population average treatment effect of treatment $j'$ relative to $j''$, defined as $\text{PATE}_{j',j''} = \mathbb{E}[Y_i(j') - Y_i(j'')]$. 
Typically, inference for the $\text{PATE}_{j',j''}$ is conducted conditional on the covariates and the treatment assignment by fitting a suitable regression model of the outcome on the treatment indicator and the covariates. 
For instance, for $g=2$, consider a linear potential outcome model of the form $Y_i(j) =  \bm{\beta}^\top_{j}\bm{B}(\bm{X}_i) + \epsilon_{ij}$ in treatment group $j \in \{1,2\}$, where  $\bm{B}(\bm{X}_i) = \{ B_1(\bm{X}_i),...,B_b(\bm{X}_i) \}^\top$ is a vector of $b$ basis functions. 
This amounts to fitting a linear regression model of the observed outcome $Y^{obs}_i = \mathbbm{1}(Z_i=1)Y_i(1) + \mathbbm{1}(Z_i=2)Y_i(2)$ on $\bm{B}(\bm{X}_i)$, $\mathbbm{1}(Z_i=1)$, and a full set of first-order interactions between $\bm{B}(\bm{X}_i)$ and $\mathbbm{1}(Z_i=1)$. 
The model-based estimator of $\text{PATE}_{12}$ is given by $\widehat{\text{PATE}}_{12} = \hat{\mathbb{E}}[Y_i(1)] - \hat{\mathbb{E}}[Y_i(2)] = \hat{\bm{\beta}}^\top_{1}\overline{\bm{B}(\bm{X})} - \hat{\bm{\beta}}^\top_{2}\overline{\bm{B}(\bm{X})}$, where $\overline{\bm{B}(\bm{X})} = \frac{1}{N}\sum_{i=1}^{N}\bm{B}(\bm{X}_i)$ and $\hat{\bm{\beta}}_z$ is the OLS estimator of $\bm{\beta}_z$. The corresponding model-based variance of the estimator of $\text{PATE}_{1,2}$ is  $Var(\widehat{\text{PATE}}_{1,2}) =  \overline{\bm{B}(\bm{X})}^\top \{Var(\hat{\bm{\beta}}_{1}) + Var(\hat{\bm{\beta}}_{2}) \}\overline{\bm{B}(\bm{X})} $. If the posited potential outcome models are true, then this conditional variance expression is valid irrespective of the underlying assignment mechanism. 
Finally, given the point estimate and its variance, we can obtain a Wald-type 95\% confidence interval for $\text{PATE}_{j',j''}$ by a Normal approximation. 
We illustrate this model-based approach and provide the associated \texttt{R} code in Section \ref{sec_inf2}.

\section[A step-by-step guide to the FSM in R]{A step-by-step guide to the FSM in \texttt{R}}
\label{sec_stepbystep}

In this section we illustrate how to use the FSM in \texttt{R}.\footnote{All computations are done under \texttt{R} version 4.0.3.}
For this we use the entire Lalonde data set.
In Section \ref{sec_example}, we give a brief description of the data and the assignment problem. 
Subsequently, in Section \ref{sec_buildup} we demonstrate how to run the FSM by choosing a selection order matrix (SOM) and a selection function.
Finally, in sections \ref{sec_designassess2} and \ref{sec_inf2}, we show how to assess the assignment and conduct model-based inference after allocation with the FSM. 

\subsection[Loading the FSM package]{Loading the \texttt{FSM} package}
\label{sec_example}

In this section, we load the \texttt{FSM} package and 
consider the complete Lalonde data set with $N = 445$ units, which we are to assign to $g=2$ groups of essentially equal size, $n_1 = 222$ and $n_2 = 223$. The data set is included in the \texttt{FSM} package. In the following code, we display the first 6 rows of the data set. 

\singlespacing
\begin{verbatim}
R> # Load the package.
R> library(FSM)
R> # Display the Lalonde dataset.
R> head(Lalonde)
\end{verbatim}

\begin{verbatim}
  Index Age Education Black Hispanic Married Nodegree Re74 Re75
1     1  17         7     1        0       0        1    0    0
2     2  17        10     1        0       0        1    0    0
3     3  17        10     1        0       0        1    0    0
4     4  17         8     1        0       0        1    0    0
5     5  17         8     1        0       0        1    0    0
6     6  17         9     1        0       0        1    0    0
\end{verbatim}
\doublespacing

In addition to the original eight covariates, we include two indicator variables \texttt{E74} and \texttt{E75} indicating whether \texttt{Re74} and \texttt{Re75} are positive, respectively. 

\singlespacing
\begin{verbatim}
R> # Include indicators for Re74 and Re75.
R> df_sample = data.frame(Lalonde, E74 = ifelse(Lalonde$Re74, 1, 0), 
+   E75 = ifelse(Lalonde$Re75, 1, 0))
\end{verbatim}
\doublespacing
Here \texttt{df\_sample} is the data frame corresponding to the Lalonde data set with a full set of $k=10$ covariates. The full-sample averages of these ten covariates are shown below.

\singlespacing
\begin{verbatim}
R> round(colMeans(df_sample), 2)
\end{verbatim}

\begin{verbatim}
    Index       Age Education     Black  Hispanic   Married  Nodegree 
   223.00     25.37     10.20      0.83      0.09      0.17      0.78 
     Re74      Re75       E74       E75 
  2102.27   1377.14      0.27      0.35 
\end{verbatim}
\doublespacing


\subsection{Running the FSM}
\label{sec_buildup}

\subsubsection{Choosing an SOM}
\label{sec_choosesom}

For the given assignment problem, we use the \texttt{som} function to create a Selection Order Matrix. Here, randomness in the SOM is governed by the SCOMARS algorithm with marginal probabilities of $T_2$ selecting at each stage equal to $\frac{n_2}{N} \approx 0.5$. The following \texttt{R} code is used to generate a particular instance of the SOM. 

\singlespacing
\begin{verbatim}
R> set.seed(7) 
R> # Specify the full sample size.
R> N = nrow(df_sample) 
R> # Specify size of T_1.
R> n1 = 222 
R> # Specify size of T_2.
R> n2 = 223 
R> # Generate a Selection Order Matrix.
R> som_obs = som(n_treat = 2, treat_sizes = c(n1, n2), method = `SCOMARS',
+   marginal_treat = rep((n2/N), N))
\end{verbatim}
\doublespacing
Here \texttt{n\_treat} indicates the number of treatment groups $g$ and \texttt{treat\_sizes} indicates the vector of group sizes, where the $j$th element corresponds to $n_j$ ($j \in \{1,2,...,g\}$). Since the SCOMARS algorithm is used to generate the SOM, we set \texttt{method = `SCOMARS'}. Note that SCOMARS is only applicable when $g=2$. For $g>2$, one can use other methods which may leverage the symmetry of the problem (see Section \ref{sec_som} and the Appendix for more details). For applying SCOMARS one needs to specify the marginal probability that $T_2$ gets to choose at the $j$th stage for all $j \in \{1,2,...,n\}$. \texttt{marginal\_treat} incorporates this vector of marginal probabilities. We display the first 10 rows of the realization of the SOM below.

\singlespacing
\begin{verbatim}
R> # Display first 10 rows of the SOM.
R> som_obs[1:10,]
\end{verbatim}
\begin{verbatim}
    Probs Treat
1  0.5011     1
2  1.0000     2
3  0.5023     2
4  0.0089     1
5  0.5034     2
6  0.0133     1
7  0.5045     1
8  1.0000     2
9  0.5057     2
10 0.0220     1
\end{verbatim}
\doublespacing

\subsubsection{Choosing a selection function and an assignment}
\label{sec_chooseself}

Given an SOM, the final assignment of the units to treatment groups is obtained using the \texttt{fsm} function of the package. The \texttt{fsm} function uses the covariate data, an SOM, and a selection function to yield the final assignment of each unit to one of the treatment groups.  The following code can be used to generate one such assignment. 

\singlespacing
\begin{verbatim}
R> # Generate a treatment assignment given som_obs.
R> f = fsm(data_frame = df_sample, SOM = som_obs, s_function = `Dopt',
+    eps = 0.0001, units_print = FALSE)
\end{verbatim}
\doublespacing

Unlike the \texttt{som} function, the \texttt{fsm} function requires access to the full data frame \texttt{df\_sample}, since it uses the covariate information from each row of the data frame to execute the selection of units. In particular, the columns of the data frame should include vectors of the transformations of the covariate vector (e.g., the original covariates, two-way interactions of the covariates, etc.) that we wish to include in a linear model of the potential outcomes. For this example, we consider a linear model of each potential outcome on the $k=10$ covariates. Also, at every stage of selection, the choosing treatment groups uses the D-optimal selection function to select the optimal unit, which is encoded by \texttt{s\_function = `Dopt'}. See the Appendix for details on the other arguments of the \texttt{fsm} function.

The output of the \texttt{fsm} function (stored in \texttt{f}) contains several objects as a list. For instance, we can generate a data frame that augments the selection order matrix with the index and the corresponding covariate vector of the selected unit at each stage using \texttt{f\$som\_appended}. The code is shown below.

\singlespacing
\begin{verbatim}
R> # Generate augmented SOM. 
R> som_obs_augmented = f$som_appended
R> # Display first 10 rows and first 10 columns of the augmented SOM.
R> round(som_obs_augmented[1:10,1:10])

\end{verbatim}
\begin{verbatim}
   Treat Index Age Education Black Hispanic Married Nodegree  Re74  Re75
1      1   387  33        11     1        0       1        1 14661 25142
2      2   147  21         8     1        0       0        1 39571  6608
3      2   176  22        10     0        0       1        1 25721 23032
4      1   144  21         7     1        0       0        1 33800     0
5      2   445  55         3     1        0       0        1     0     0
6      1   327  28        11     0        1       1        1  3473     0
7      1   187  23         8     0        0       1        1     0  1713
8      2   301  27        13     0        0       1        0  9382   854
9      2   443  50        10     0        1       0        1     0     0
10     1   442  48         4     1        0       0        1     0     0
\end{verbatim}
\doublespacing

Alternatively, a data frame with an additional column of the treatment indicator can be obtained using \texttt{f\$data\_frame\_allocated}. In the following code, we extract the vector of treatment labels (ordered by the unit indices) from this augmented data frame.

\singlespacing
\begin{verbatim}
R> # Augment df_sample with the treatment label.
R> df_sample_aug = f$data_frame_allocated
R> # Create a vector of observed treatment labels.
R> Z_fsm_obs = df_sample_aug$Treat
\end{verbatim}
\doublespacing

\subsection{Assessing the design}
\label{sec_designassess2}

To assess covariate balance under the given realization of the FSM, we draw Love plots \citep{love2004graphical} of the TASMDs of the covariates using the \texttt{love\_plot} function of the package. For comparison, we also consider a random realization of CRD, drawn using the \texttt{crd} function of the package. 

\singlespacing
\begin{verbatim}
R> # Generate an assignment under CRD.
R> Z_crd = crd(df_sample, n_treat = 2, treat_sizes = c(n1, n2), 
control = FALSE)$Treat

R> # Generate Love plot of the TASMDs in T_1. 
R> love_plot(data_frame = df_sample, index_col = TRUE, 
+   alloc1 = Z_fsm_obs, alloc2 = Z_crd, imbalance = 'TASMD', treat_lab = 1, 
+   legend_text = c('FSM','CRD'), xupper =  0.15) 
\end{verbatim}
\doublespacing

In the \texttt{love\_plot} function, \texttt{data\_frame} includes a data frame containing the transformations of the covariate vector whose TASMDs we want to compute (along with an optional column for the index of the units). \texttt{alloc1} and \texttt{alloc2} are inputs for the two vectors of treatment assignments which, in this case, are the realized assignment vectors under the FSM and CRD, respectively. \texttt{imbalance} indicates the measure of imbalance used, which can be either \texttt{`TASMD'} or \texttt{`ASMD'}, the latter corresponding to the absolute standardized mean differences (ASMD). 
The argument \texttt{treat\_lab = 1} indicates that the TASMD is computed in $T_1$. Since the two group sizes are roughly equal, the corresponding plot of the TASMDs in $T_2$ is similar. 
See the package documentation for more details on the other arguments. 

\begin{figure}[!ht]
\centering
\begin{subfigure}{.5\textwidth}
  \centering
  \includegraphics[width=\linewidth]{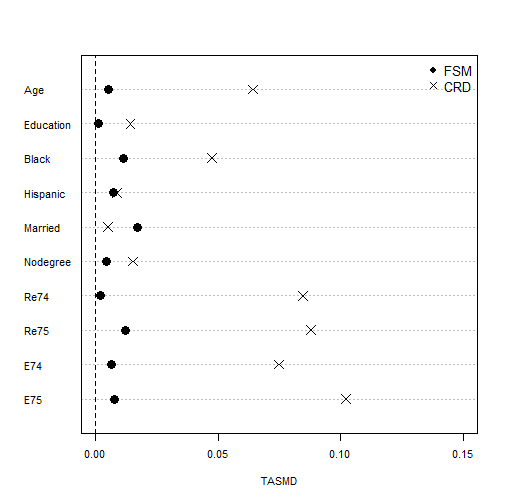}
  \label{fig:sub1}
\end{subfigure}%
\begin{subfigure}{.5\textwidth}
  \centering
  \includegraphics[width=\linewidth]{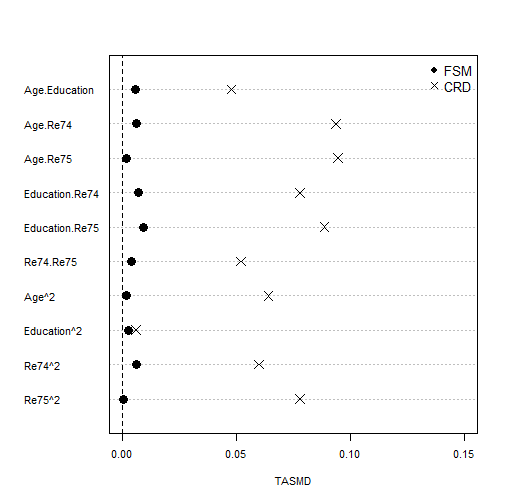}
  \label{fig:sub2}
\end{subfigure}
\caption{\footnotesize Love plot of the TASMDs of the original covariates (left) and the squares and pairwise products of \texttt{Age}, \texttt{Education}, \texttt{Re74}, and \texttt{Re75} (right) in treatment group 1 for the given realizations of CRD and the FSM.}
\label{fig:loveplots}
\end{figure}

To check balance on other transformations of the covariate vector, we also generate Love plots of the TASMDs of the squares of \texttt{Age}, \texttt{Education}, \texttt{Re74} and \texttt{Re75}, and their pairwise products. See the Appendix for the associated code.
The resulting two plots are shown in Figure \ref{fig:loveplots}. 
It is evident that the given realization of the FSM yields better overall mean-balance on the observed covariates and their transformations in $T_1$ (relative to the full sample) than that of CRD, as the corresponding TASMDs under the realized FSM are almost uniformly smaller than those under the realized CRD. Thus, Figure \ref{fig:loveplots} reiterates the balancing property of the FSM under both correct and incorrect specifications of the underlying potential outcome models, as discussed in Section \ref{sec_designassess1}. 

Indeed, Figure \ref{fig:loveplots} compares covariate balance between specific realizations of CRD and the FSM and hence is not fully reflective of their comparative performances across repeated randomizations. To address this, for each design we obtain the distribution of the TASMDs in $T_1$ across all the 10 covariates and across 1000 independent realizations of the design. To calculate the TASMDs for each realization of CRD and the FSM, we use the \texttt{tasmd\_rand} function of the package. A typical use of the \texttt{tasmd\_rand} function for a given assignment vector under CRD (denoted by \texttt{Z\_crd}) and that under the FSM (denoted by \texttt{Z\_fsm}) is shown below.

\singlespacing
\begin{verbatim}
R> tasmd_rand(data_frame = df_sample, index_col = TRUE, 
+   alloc1 = Z_crd, alloc2 = Z_fsm, treat_lab = 1, legend = c(`CRD', `FSM'))
\end{verbatim}
\doublespacing

\begin{figure}[H]
    \centering
    \includegraphics[height= 7.5cm, width =  0.6\textwidth]{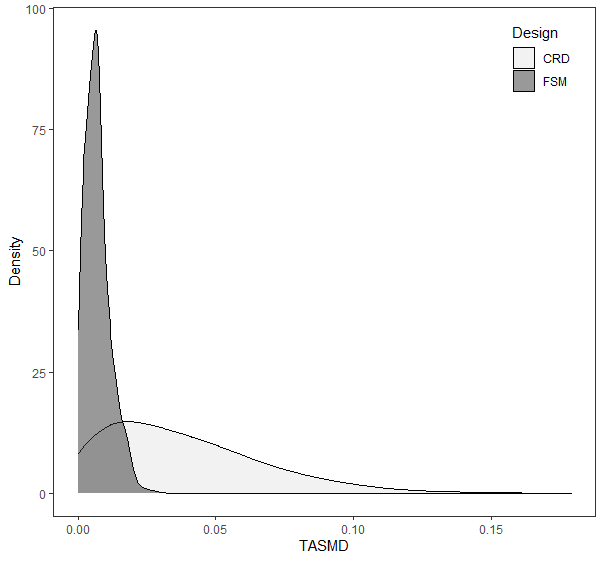}
    \caption{\footnotesize Distribution of the TASMDs across covariates and 1000 independent realizations of CRD and the FSM.}
    \label{fig:tasmd_density}
\end{figure}

The arguments of \texttt{tasmd\_rand} are the same as those in the \texttt{love\_plot} function, barring a few exceptions (please see the Appendix and the package documentation for details). Using the output from this function, in Figure \ref{fig:tasmd_density} we provide density plots of the distributions of the TASMDs under CRD and the FSM across all the covariates and 1000 realization of the respective assignment mechanisms.

Figure \ref{fig:tasmd_density} shows that the randomization distribution of the TASMDs under the FSM are substantially more concentrated towards zero than those under CRD. In fact, the mean and standard deviation of the TASMDs under the FSM are less than $19\%$ and $17\%$ of those under CRD, respectively. Moreover, for several realizations of the CRD the resulting TASMDs are larger than 0.15 (which translates to ASMDs larger than 0.3), indicating assignments that are highly imbalanced on one or more covariates. Such extreme assignments are ruled out by design under the FSM.

Finally, we generate the distribution of the model-based ESS of the realized assignment under the FSM relative to CRD. Here we simulate the potential outcomes from the model $Y(1) = 100 - \texttt{Age} + 6 \times \texttt{Education} - 20 \times \texttt{Black} + 20 \times \texttt{Hispanic} + 0.003 \times \texttt{Re75} + \epsilon$, $Y(2) = Y(1)$; where $\epsilon$ is a Normal random variable with mean 0, variance 16.

\singlespacing
\begin{verbatim}
R> set.seed(9)
R> # Generate the potential outcomes Y_1 and Y_2.
R> Y_1 = 100 - df_sample$Age + 6 * df_sample$Education - 20 * df_sample$Black + 
+   20 * df_sample$Hispanic + 0.003 * df_sample$Re75 + rnorm(N, 0, 4)
R> Y_1 = round(Y_1, 2)
R> # Set the unit level causal effect tau as zero.
R> tau = 0
R> Y_2 = Y_1 + tau0
R> # Create a matrix of potential outcomes.
R> Y_appended = cbind(Y_1, Y_2)
\end{verbatim}
\doublespacing

Using the \texttt{ess\_model} function, we compute the vector of the ESS for the realized FSM relative to 1000 realizations of CRD generated earlier. A typical use of the \texttt{ess\_model} function for a given assignment vector under CRD and that under the FSM is shown below.

\singlespacing
\begin{verbatim}
R>  ess_model(X_cov = df_sample[,-1], assign_matrix = cbind(Z_crd, Z_fsm), 
+   Y_mat = Y_appended, contrast = c(1,-1))
\end{verbatim}
\doublespacing

The \texttt{ess\_model} function requires the matrix of covariates or transformations thereof (denoted by \texttt{X\_cov}) that will be used as explanatory variables in the linear outcome models within each treatment group. Also, it requires the matrices of treatment assignments (denoted by \texttt{assign\_matrix}) and the potential outcomes (denoted by \texttt{Y\_mat}). Finally, the coefficients of the treatment contrast of interest is included as a vector (denoted by \texttt{contrast}). Here we focus on estimating $\text{PATE}_{1,2} = \mathbb{E}[Y_i(1) - Y_i(2)]$, and thus we set \texttt{contrast = c(1,-1)}. See the Appendix for the exact use of this function for this example. The ESS values for CRD and the FSM are stored in \texttt{Neff\_crd} and \texttt{Neff\_fsm} respectively.

\begin{figure}[!ht]
    \centering
    \includegraphics[height = 9.5cm, width =  0.6\textwidth]{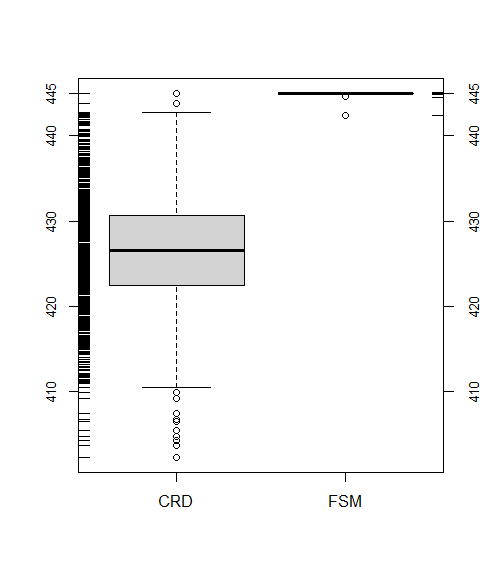}
    \caption{\footnotesize Boxplots of the distribution of model-based ESS for the given realization of the FSM with respect to 1000 independent realizations of CRD.}
    \label{fig:ess boxplot2}
\end{figure}

We now compare the ESS of the two designs using side-by-side boxplots. The resulting plot is shown in Figure \ref{fig:ess boxplot2}.

\singlespacing
\begin{verbatim}
R> Create boxplots of the ESS.
R> boxplot(Neff_crd, Neff_fsm, axes = F, names = c(`CRD', `FSM'))
R> # See the Appendix for codes for specifications of the axes marks.        
\end{verbatim}
\doublespacing

Similar to the example in Section \ref{sec_designassess1}, the distribution of the ESS for the realized FSM is substantially more concentrated towards the full-sample size of 445 than that under CRD. In particular, \texttt{Neff\_fsm} equals 445 for at least 75 percent of the iterations, as indicated by the first quartile. The improvement in the median of \texttt{Neff\_fsm} over the median of \texttt{Neff\_crd} is roughly $5 \%$ which, although less striking than the improvement noted in Section \ref{sec_designassess1}, is still meaningful in terms of allocation of resources in practice. The model-based ESS conditions on the assignment and only takes into account the variability from random sampling of the units. If we take into account the variability due to the random assignment mechanism (which is considered in randomization-based ESS), the FSM leads to considerable improvement of the ESS over CRD. See the Appendix for the corresponding codes for computing randomization-based ESS.


\subsection{Making inference with the FSM}
\label{sec_inf2}

In this section we show how to conduct model-based inference with the FSM. We refer the reader to the package documentation for details on functions to conduct randomization-based inference. Here we use the set of potential outcomes \texttt{Y\_1} and \texttt{Y\_2} generated in the previous section.
For model-based inference, we fit separate linear models of the observed outcome (denoted by \texttt{Y\_obs}) on all the covariates (with an intercept term) in the two treatment groups. 
The covariates are demeaned to ensure that the estimated treatment effect is same as the difference between the estimated coefficients of the intercept terms from the two models. The resulting matrix of covariates is denoted by \texttt{X\_cov\_demean}. 

\singlespacing
\begin{verbatim}
R> # Fit a linear model in T_1.
R> fit_t1 = lm(Y_obs[Z_fsm_obs == 1] ~ X_cov_demean[Z_fsm_obs == 1,])  
R> # Fit a linear model in T_2.
R> fit_t2 = lm(Y_obs[Z_fsm_obs == 2] ~ X_cov_demean[Z_fsm_obs == 2,])
R> # Compute point estimate of the ATE.
R> T0 = fit_t1$coefficients[1] - fit_t2$coefficients[1]
R> as.numeric(T0)

\end{verbatim}
\begin{verbatim}
[1] -0.04956801
\end{verbatim}
\doublespacing
Thus, the point estimate of the average treatment effect of treatment 1 versus treatment 2 is approximately -0.05. 
The following code is used to calculate the model-based standard error of this estimator. 

\singlespacing
\begin{verbatim}
R> # Compute the variance of the estimator.
R> V0 = (coef(summary(fit_t1))[1, `Std. Error'])^2 +
+   (coef(summary(fit_t2))[1, `Std. Error'])^2 
R> # Display the standard error of the estimator.
R> sqrt(V0)

\end{verbatim}
\begin{verbatim}
[1] 0.364697
\end{verbatim}
\doublespacing

The corresponding asymptotic 95\% confidence interval is computed using the following.

\singlespacing
\begin{verbatim}
R> # Compute the 95% Wald confidence interval.
R> CI = c(T0 - 1.96 * sqrt(V0), T0 + 1.96 * sqrt(V0))
R> names(CI) = c(`Lower', `Upper')
R> CI

\end{verbatim}
\begin{verbatim}
     Lower      Upper 
-0.7643742  0.6652382 
\end{verbatim}
\doublespacing

Therefore, the 95\% confidence interval for $\text{PATT}_{1,2}$ is approximately (-0.76, 0.67). 
We note that in this case, the interval contains the true average treatment effect of treatment 1 versus treatment 2, which equals zero. 

\section{Discussion}
\label{sec_discussion}

In this paper, we have provided a practical introduction to the FSM. Using as an empirical example the NSW experiment, we have illustrated the use of the \texttt{FSM} package for \texttt{R}. 
In the example we also illustrate the performance of the FSM in comparison to CRD, and show that the FSM tends to achieve better balance on multiple covariates (or transformations thereof) and higher effective sample sizes. 
The example pertains to the assignment problem with two treatment groups; however, the methodology directly applies to more groups.  

A number of methodological developments on the FSM are currently underway. 
Here we discuss two major areas of development. First, the selection function component of the FSM discussed in this paper is based on D-optimality. 
However, other selection functions can also be used. 
For example, in the HIE a selection function based on A-optimality criteria was used. 
A-optimal selection functions require a priori specifications of \textit{policy weights} that allow researchers to emphasize some main effects of the model as more important than others. 
We recommend using the D-optimality criteria as it satisfies several desirable statistical properties, while circumventing the specification of policy weights. 
However, there is room for further development of the selection function to handle more complex scenarios. 
For instance, the D-optimal selection function can be modified/replaced for generalized linear models of the potential outcomes with non-linear link functions. 
Also, more work needs to be done on the selection function to ensure balance on a high (possibly infinite) dimensional class of basis functions of the covariates. 

Second, the FSM can be extended to more complex experimental designs, such as sequential, stratified, and cluster randomized experiments. 
In particular, a modified version of the FSM which we call \textit{batched FSM} is being developed to assign units into treatment groups, when the units arrive sequentially in batches. 
Also, there is work in progress for devising a principled algorithm that generates a fair and random SOM in stratified experiments.

\section*{Acknowledgments}
This work was supported through a grant from the Alfred P. Sloan Foundation (G-2020-13946).

\pagebreak
\onehalfspacing
\bibliographystyle{asa}
\bibliography{mybibliography21}

\newpage

\newpage

\begin{appendix}

\section{The \texttt{som} function} \label{app:technical}

\subsection{\texttt{R} function}

\texttt{som(data\_frame = NULL, n\_treat, treat\_sizes, include\_discard = FALSE, method = `SCOMARS', control = FALSE, marginal\_treat = NULL)}

\subsection{Arguments}

\begin{itemize}
\item \texttt{data\_frame}: A (optional) data frame corresponding to the full sample of units.. Required if \texttt{include\_discard = TRUE}.

\item \texttt{n\_treat}: Number of treatment groups.

\item  \texttt{treat\_sizes}: A vector of treatment group sizes. If \texttt{control = TRUE}, the first element of \texttt{treat\_sizes} should be the control group size.

\item \texttt{include\_discard}: \texttt{TRUE} if a discard group is considered.

\item \texttt{method}: Specifies the selection strategy used among \texttt{`global percentage'}, \texttt{`randomized chunk'}, \texttt{`SCOMARS'}. \texttt{`SCOMARS'} is applicable only if \texttt{n\_treat = 2}. For multiple groups ($g>2$) of sizes $cm_1,cm_2,...,c m_g$, where $c$ is an integer and $m_1, m_2,..., m_g$ are coprime integers, \texttt{method = `randomized chunk'} creates an SOM by randomly permuting the chunk $(\underbrace{1,1,...,1}_{m_1},\underbrace{2,2,...,2}_{m_2},...,\underbrace{g,g,...,g}_{m_g})$ $c$ times independently. For multiple groups ($g>2$), \texttt{method = `global percentage'} determines the choosing treatment group as the group that has made the lowest proportion of choices (relative to its total size) upto the previous stage. Ties are resolved randomly.

\item \texttt{control}: if \texttt{TRUE}, treatments are labelled as 0,1,...,g-1 (0 representing the control group). If \texttt{FALSE}, they are labelled as 1,2,...,g.

\item \texttt{marginal\_treat}: A vector of marginal probabilities, the jth element being the probability that treatment group (or treatment group 2 in case \texttt{control = FALSE}) gets to choose at the jth stage given the total number of choices made by treatment group upto the (j-1)th stage. Only applicable when \texttt{method = `SCOMARS'}.
 
\end{itemize}

\subsection{Value}
A data frame containing the selection order of treatments, i.e. the labels of treatment groups at each stage of selection. If \texttt{method = `SCOMARS'}, the data frame contains an additional column of the conditional selection probabilities.

\section{The \texttt{fsm} function}
\label{sec_fsmfunc}

\subsection{\texttt{R} function}

    \texttt{fsm(data\_frame, SOM, s\_function = `Dopt', Q\_initial = NULL, eps = 0.001, ties = `random', intercept = TRUE, standardize = TRUE, units\_print = TRUE, index\_col = TRUE, Pol\_mat = NULL, w\_pol = NULL)}

\subsection{Arguments}

\begin{itemize}
\item \texttt{data\_frame}: A data frame containing a column of unit indices (optional) and covariates (or transformations thereof).

\item  \texttt{SOM}: A selection order matrix.

\item \texttt{s\_function}: Specifies a selection function, a string among \texttt{`constant'}, \texttt{`Dopt'},  \texttt{`Aopt'}, \texttt{`max pc'}, \texttt{`min pc'}, \texttt{`Dopt pc'}, \texttt{`max average'}, \texttt{`min average'}, \texttt{`Dopt average'}. \texttt{`constant'} selection function puts a constant value on every unselected unit. \texttt{`Dopt'} use the D-optimality criteria based on the full set of covariates to select units. \texttt{`Aopt'} uses the A-optimality criteria. \texttt{`max pc'} (respectively, \texttt{`min pc'}) selects that unit that has the maximum (respectively, minimum) value of the first principal component. \texttt{`Dopt pc'} uses the D-optimality criteria on the first principal component, \texttt{`max average'} (respectively, \texttt{`min average'}) selects that unit that has the maximum (respectively, minimum) value of the simple average of the covariates. \texttt{`Dopt average'} uses the D-optimality criteria on the simple average of the covariates.

\item  \texttt{Q\_initial}: A (optional) non-singular matrix (called `initial matrix') that is added the $(X^T X)$ matrix of the choosing treatment group at any stage, when the $(X^T X)$ matrix of that treatment group at that stage is non-invertible. If \texttt{FALSE}, the $(X^T X)$ matrix for the full set of observations is used as the non-singular matrix. Applicable if \texttt{s\_function = `Dopt'} or \texttt{`Aopt'}.

\item \texttt{eps}: Proportionality constant for \texttt{Q\_initial}, the default value is 0.001.

\item \texttt{ties}: Specifies how to deal with ties in the values of the selection function. If \texttt{ties = `random'}, a unit is selected randomly from the set of candidate units. If \texttt{ties = `smallest'}, the unit that appears earlier in the data frame, i.e. the unit with the smallest index gets selected.

\item \texttt{intercept}: if \texttt{TRUE}, the design matrix (within each treatment arm) includes a column of intercepts.

\item \texttt{standardize}: if \texttt{TRUE}, the columns of the $\underline{\bm{X}}$ matrix other than the column for the intercept (if any), are standardized.

\item \texttt{units\_print}: if \texttt{TRUE}, the function automatically prints the candidate units at each step of the build-up process.

\item \texttt{index\_col}: if \texttt{TRUE}, \texttt{data\_frame} contains a column of unit indices.

\item \texttt{Pol\_mat}: Applicable only when \texttt{s\_function = `Aopt'}.

\item \texttt{w\_pol}: A vector of policy weights. Applicable only when \texttt{s\_function = `Aopt'}.

\end{itemize}

\subsection{ Value}

The function returns a list containing the following items.

\begin{itemize}
\item \texttt{data\_frame\_allocated}:  The original data frame augmented with the column of the treatment indicator.
\item \texttt{som\_appended}:  The SOM with augmented columns for the indices and covariate values for units selected.
\item  \texttt{som\_split}:  \texttt{som\_appended}, split by the levels of the treatment.
\item \texttt{crit\_print}:  The value of the objective function, at each stage of build up process. At each stage, the unit that maximizes the objective function is selected.
\end{itemize}

\section{Replication code for Section 3}

\singlespacing
\begin{verbatim}
R> # Load the required packages.
R> library(ggplot2)
R> library(ggthemes)

R> # Load the package.
R> library(FSM)
R> # Display the Lalonde dataset.
R> head(Lalonde)
R> # Include indicators for Re74 and Re75.
R> df_sample = data.frame(Lalonde, E74 = ifelse(Lalonde$Re74, 1, 0), 
+   E75 = ifelse(Lalonde$Re75, 1, 0))
R> round(colMeans(df_sample), 2)

R> set.seed(7) 
R> # Specify the full sample size.
R> N = nrow(df_sample) 
R> # Specify size of T_1.
R> n1 = 222 
R> # Specify size of T_2.
R> n2 = 223 
R> # Generate a Selection Order Matrix.
R> som_obs = som(data_frame = NULL, n_treat = 2, treat_sizes = c(n1, n2),
+   method = 'SCOMARS', marginal_treat = rep((n2/N), N))
R> # Display first 10 rows of the SOM.
R> som_obs[1:10,]

R> # Generate a treatment assignment given som_obs.
R> f = fsm(data_frame = df_sample, SOM = som_obs, s_function = 'Dopt',
+    eps = 0.0001, units_print = FALSE)

R> # Generate augmented SOM. 
R> som_obs_augmented = f$som_appended
R> # Display first 10 rows and first 10 columns of the augmented SOM.
R> round(som_obs_augmented[1:10,1:10])
R> # Augment df_sample with the treatment label.
R> df_sample_aug = f$data_frame_allocated
R> # Create a vector of observed treatment labels.
R> Z_fsm_obs = df_sample_aug$Treat

R> # Generate an assignment under CRD.
R> Z_crd = crd(df_sample, n_treat = 2, treat_sizes = c(n1, n2), 
control = FALSE)$Treat

R> # Generate Love plot of the TASMDs in T_1. 
R> love_plot(data_frame = df_sample, index_col = TRUE, 
+   alloc1 = Z_fsm_obs, alloc2 = Z_crd, imbalance = 'TASMD', treat_lab = 1, 
+   legend_text = c('FSM','CRD'), xupper =  0.15)

R> # Check balance on squares and products of continuous covariates
R> df_sample_sq_int  = make_sq_inter(data_frame = df_sample[,c(2,3,8,9)], 
+   is_square = TRUE, is_inter = TRUE, keep_marginal = FALSE)
R> love_plot(data_frame = df_sample_sq_int,  index_col = FALSE,
+   alloc1 = Z_fsm_obs, alloc2 = Z_crd, imbalance = "TASMD", treat_lab = 1, 
+   legend_text = c('FSM','CRD'), xupper = 0.15) 

R> # Generate 1000 indep realizations of CRD & the FSM
R> set.seed(7)
R> # Set number of iterations
R> n_iter = 1000
R> # Initialize assignment vectors for all the iterations
R> Z_crd_iter = matrix(rep(0,n_iter*N),nrow = n_iter)
R> Z_fsm_iter = matrix(rep(0,n_iter*N),nrow = n_iter)
R> for(i in 1:n_iter)
R> {
R> # Generate assignment under CRD
R> fc = crd(df_sample, n_treat = 2, treat_sizes = c(n1,n2), control = FALSE)
R> Z_crd_iter[i,] = fc$Treat

R> # Generate assignment under the FSM
R>  som_iter = som(data_frame = NULL, n_treat = 2, treat_sizes = c(n1,n2),
+   method = 'SCOMARS', marginal_treat = rep((n2/N),N))
R>  f = fsm(data_frame = df_sample, SOM = som_iter, s_function = 'Dopt',
+   eps = 0.0001, units_print = FALSE)
R>  Z_fsm_iter[i,] = f$data_frame_allocated$Treat
R> }

R> # Compute the distribution of TASMD across different randomizations and
+   across all the covariates
R> # Compute the number of original covariates
R> k = ncol(df_sample)-1
R> # Initialize vectors of TASMDs for both designs
R> TASMD_crd_iter = matrix(rep(0, n_iter*k), nrow = n_iter)
R> TASMD_fsm_iter = matrix(rep(0, n_iter*k), nrow = n_iter)
R> for(r in 1:n_iter)
R> {
R>  TASMD_iter = tasmd_rand(data_frame = df_sample, index_col = TRUE, 
+   alloc1 = Z_crd_iter[r,], alloc2 = Z_fsm_iter[r,], 
+   treat_lab = 1, legend = c('CRD','FSM'), roundoff = 3)
R>  TASMD_crd_iter[r,] = TASMD_iter[,1]
R>  TASMD_fsm_iter[r,] = TASMD_iter[,2]
R> }
R> #  Concatenate the TASMDs across all the covariates
R> TASMD_crd_concat = as.vector(TASMD_crd_iter)
R> TASMD_fsm_concat = as.vector(TASMD_fsm_iter)

R> # Plot estimated densities of the TASMDs under CRD and the FSM
R> TASMD_df = data.frame(TASMD = c(TASMD_crd_concat, TASMD_fsm_concat),
+   Design = rep(c('CRD', 'FSM'), each = length(TASMD_crd_concat)))
R> p = ggplot(TASMD_df, aes(x = TASMD, fill = Design)) + 
+   geom_density(alpha = 0.5, adjust = 2.5)
R> p + labs(y = 'Density') + theme_bw() + 
+   theme(panel.grid.major = element_blank(), panel.grid.minor = element_blank())+
+   scale_fill_grey(start = 0.9, end = .2) + theme(legend.position = c(0.9, 0.9))

R> set.seed(9)
R> # Generate the potential outcomes Y_1 and Y_2.
R> Y_1 = 100 - df_sample$Age + 6 * df_sample$Education - 20 * df_sample$Black + 
+   20 * df_sample$Hispanic + 0.003 * df_sample$Re75 + rnorm(N, 0, 4)
R> Y_1 = round(Y_1, 2)
R> # Set the unit level causal effect tau as zero.
R> tau = 0
R> Y_2 = Y_1 + tau
R> # Create a matrix of potential outcomes.
R> Y_appended = cbind(Y_1, Y_2)
R> # Create matrix of observed covariates
R> X_cov = df_sample[,-1]

R> # Initialize model-based ESS for CRD and the FSM
R> Neff_crd = rep(0,n_iter)
R> Neff_fsm = rep(0,n_iter)
R> for(r in 1:n_iter)
R> {
R>  Z_crd_obs = Z_crd_iter[r,]
R>  # Create matrix of assignments
R>  Z_big = cbind(Z_crd_obs, Z_fsm_obs)
R>  # Calculate the ESS  
R>  ess = ess_model(X_cov = X_cov, assign_matrix = Z_big, 
+   Y_mat = Y_appended, contrast = c(1, -1))
R>  Neff_crd[r] = ess[1]
R>  Neff_fsm[r] = ess[2]
R> }

R> # Draw boxplots of the ESS
R> boxplot(Neff_crd, Neff_fsm, axes = F, names = c('CRD', 'FSM'))
R> # Set axes marks
R> axis(1, at = c(1:2), labels = c('CRD', 'FSM'))
R> axis(2, cex.axis = 0.8, at = c(seq(390,450,10),445))
R> axis(4, cex.axis = 0.8, at = c(seq(390,450,10),445))
R> rug(x =jitter(Neff_crd), side = 2)
R> rug(x =jitter(Neff_fsm), side = 4)
R> box()

R> # Compute randomization-based ESS.
R> # Construct 3-dim array of the assignments.
R> # 1st coordinate represents iterations.
R> # 2nd coordinate represents units.
R> # 3rd coordinate represents designs.
R> Z_array = array(0, dim = c(n_iter, N, 2))
R> Z_array[,, 1] = Z_crd_iter
R> Z_array[,, 2] = Z_fsm_iter
R> # Calculate ESS of CRD vs the FSM
R> ess_rand(assign_array = Z_array, Y_mat = Y_appended, contrast = c(1, -1))

R> # Create a vector of observed outcome
R> Y_obs = Y_1
R> Y_obs[Z_fsm_obs == 2] = Y_2[Z_fsm_obs == 2]
R> # De-mean columns of X_cov
R> X_cov_demean = apply(X_cov, 2, function(y) y - mean(y))

R> # Fit a linear model in T_1.
R> fit_t1 = lm(Y_obs[Z_fsm_obs == 1] ~ X_cov_demean[Z_fsm_obs == 1,])  
R> # Fit a linear model in T_2.
R> fit_t2 = lm(Y_obs[Z_fsm_obs == 2] ~ X_cov_demean[Z_fsm_obs == 2,])
R> # Compute point estimate of the ATE.
R> T0 = fit_t1$coefficients[1] - fit_t2$coefficients[1]
R> as.numeric(T0)

R> # Compute the variance of the estimator.
R> V0 = (coef(summary(fit_t1))[1, 'Std. Error'])^2 +
+   (coef(summary(fit_t2))[1, 'Std. Error'])^2 
R> # Display the standard error of the estimator.
R> sqrt(V0)

R> # Compute the 95% Wald confidence interval.
R> CI = c(T0 - 1.96 * sqrt(V0), T0 + 1.96 * sqrt(V0))
R> names(CI) = c('Lower', 'Upper')
R> CI

\end{verbatim}
\doublespacing

\end{appendix}


\end{document}